\documentclass{article}
\PassOptionsToPackage{numbers, square, compress}{natbib}
\usepackage[final]{neurips_2020_ml4ps}
\usepackage[utf8]{inputenc} % allow utf-8 input
\usepackage[T1]{fontenc}    % use 8-bit T1 fonts
\usepackage{hyperref}       % hyperlinks
\hypersetup{
    colorlinks=true,
    linkcolor=cyan,
    filecolor=magenta,      
    urlcolor=blue,
    citecolor=blue,
}
\usepackage{url}            % simple URL typesetting
\usepackage{booktabs}       % professional-quality tables
\usepackage{amsfonts,amsmath}       % blackboard math symbols
\usepackage{nicefrac}       % compact symbols for 1/2, etc.
\usepackage{microtype}      % microtypography
\usepackage{xcolor}
\usepackage{graphicx}

\title{Accelerated Charged Particle Tracking with Graph Neural Networks on FPGAs}

\usepackage{xspace}

\newcommand{\hlsfml}{\texttt{hls4ml}\xspace}

\newcommand{\pt}{\ensuremath{p_{\mathrm{T}}}\xspace}

\author{
  Aneesh Heintz\thanks{These two authors contributed equally.} \\
  Cornell Unviersity \\
  Ithaca, NY 14850, USA\\
  \And
  Vesal Razavimaleki*, Javier Duarte\\
  University of California San Diego\\
  La Jolla, CA 92093, USA \\
  \And
  Gage DeZoort, Isobel Ojalvo, Savannah Thais \\
  Princeton University \\
  Princeton, NJ 08544, USA \\
  \And
  Markus Atkinson, Mark Neubauer\\
  University of Illinois at Urbana-Champaign \\
  Champaign, IL 61820, USA\\
  \And
  Lindsey Gray, Sergo Jindariani, Nhan Tran \\
  Fermi National Accelerator Laboratory \\
  Batavia, IL 60310, USA\\
  \And
  Philip Harris, Dylan Rankin \\
  Massachusetts Institute of Technology \\
  Cambridge, MA, 02139, USA\\
  \And
  Thea Aarrestad, Vladimir Loncar\thanks{Also at Institute of Physics Belgrade, Belgrad, Serbia.}, Maurizio Pierini, Sioni Summers \\
  European Organization for Nuclear Research (CERN) \\
  CH-1211 Geneva 23, Switzerland
  \And
  Jennifer Ngadiuba \\
  California Institute of Technology \\
  Pasadena, CA 92115, USA
  \And
  Mia Liu \\
  Purdue University \\
  West Lafayette, IN 47907, USA
  \And
  Edward Kreinar\\
  HawkEye360\\
  Herndon, VA 20170, USA
  \And
  Zhenbin Wu\\
  University of Illinois at Chicago\\
  Chicago, IL 60607, USA
}

\begin{document}

\begin{flushright}
FERMILAB-CONF-20-622-CMS-SCD
\end{flushright}

\maketitle

\begin{abstract}
We develop and study FPGA implementations of algorithms for charged particle tracking based on graph neural networks. 
The two complementary FPGA designs are based on OpenCL, a framework for writing programs that execute across heterogeneous platforms, and \hlsfml, a high-level-synthesis-based compiler for neural network to firmware conversion.
We evaluate and compare the resource usage, latency, and tracking performance of our implementations based on a benchmark dataset.
We find a considerable speedup over CPU-based execution is possible, potentially enabling such algorithms to be used effectively in future computing workflows and the FPGA-based Level-1 trigger at the CERN Large Hadron Collider.
\end{abstract}

\section{Introduction}

In high energy physics (HEP), charged particle tracking~\cite{Amrouche:2019wmx,Strandlie:2010zz} is a crucial task necessary for the accurate determination of the kinematics of the particles produced in a collision event.
%, including the position, direction, and momentum of the particles at their production points. 
%This task leverages specialized detectors positioned close to the beam collision area in a strong magnetic field.
%Charged particles created in the collisions bend in the magnetic field and ionize the material of these detectors, providing position measurements along the trajectory of each particle. 
The objective of tracking algorithms is to identify the trajectories of charged particles created in the collisions that bend in a magnetic field and ionize the material of detectors, providing position measurements along the trajectory of each particle.
Current tracking algorithms~\cite{Chatrchyan:2014fea,Aaboud:2017all,combkalman1,combkalman2,combkalman3,kalman} scale worse than quadratically in the number of hits, which are expected to increase dramatically at higher beam intensities. % and for more granular sensitive detectors.
This motivates the study of alternative algorithms with different computational scaling. 
Another important consideration is the ability to accelerate these algorithms using highly-parallel heterogeneous computing resources like graphics processing units (GPUs) and field-programmable gate arrays (FPGAs) as further improvements in single-core CPU performance may be limited~\cite{breakdown,dennard}.
Recent efforts~\cite{Farrell:2018cjr,Ju:2020xty} have demonstrated the effectiveness of graph neural networks (GNNs) to correctly classify ``segments'' belonging to tracks.
Graph-based approaches are well suited to this task because tracking data can be naturally encoded as a graph structure~\cite{Shlomi:2020gdn} and GNNs consider local information between pairs of hits to learn relationships between them in order to ``connect the dots'' to infer tracks.
%In other words, track finding is an example of edge classification on a graph data structure.

In this paper, we accelerate a GNN~\cite{Ju:2020xty} for segment classification, based on the interaction network (IN) architecture~\cite{Battaglia:2016jem,battaglia2018relational}, with FPGA implementations. 
Such implementations enable efficient processing, in both speed and energy consumption for large HEP datasets.
%being collected in HEP experiments now and in the future. 
They may also enable the use of GNNs in the high-throughput, FPGA-based data filter system, known as the Level-1 trigger~\cite{ATLASL1T,ATLASP2L1T,CMSL1T,CMSP2L1T}, which has strict sub-microsecond latency requirements that only FPGAs or application-specific integrated circuits (ASICs) can meet.
%In Section~\ref{sec:in}, we describe the structure of the IN model.
We design two complementary implementations using OpenCL~\cite{OpenCL} and \hlsfml~\cite{Duarte:2018ite,vloncar_2020_4161550}, a specialized compiler for converting machine learning (ML) algorithms into FPGA firmware.
We evaluate the resource usage, latency, and tracking performance of our implementations based on the benchmark TrackML dataset~\cite{Amrouche:2019wmx}.
%Finally, we summarize our findings in Section~\ref{sec:summary}.

\section{TrackML Data and Interaction Network Models}
\label{sec:in}

To benchmark the algorithms, we use the TrackML dataset~\cite{Amrouche:2019wmx}, which consists of simulated 3D measurements of particles coming from independent collision events at the LHC.
%The training dataset contains the recorded hits, their ground truth counterpart and their association to particles, and the initial parameters of those particles, while the test dataset contains only the recorded hits.
For our application, the data is embedded as a graph by considering the hits as nodes and pairs of hits on adjacent layers as edges.
The edges corresponding to true track segments are labeled as one and all others are labeled as zero.
The goal of the segment classifier algorithm is to correctly classify the edges as true or spurious.

\begin{figure}[t]
    \centering
    \includegraphics[width=0.33\textwidth,clip=true,viewport = 0 10 392 382]{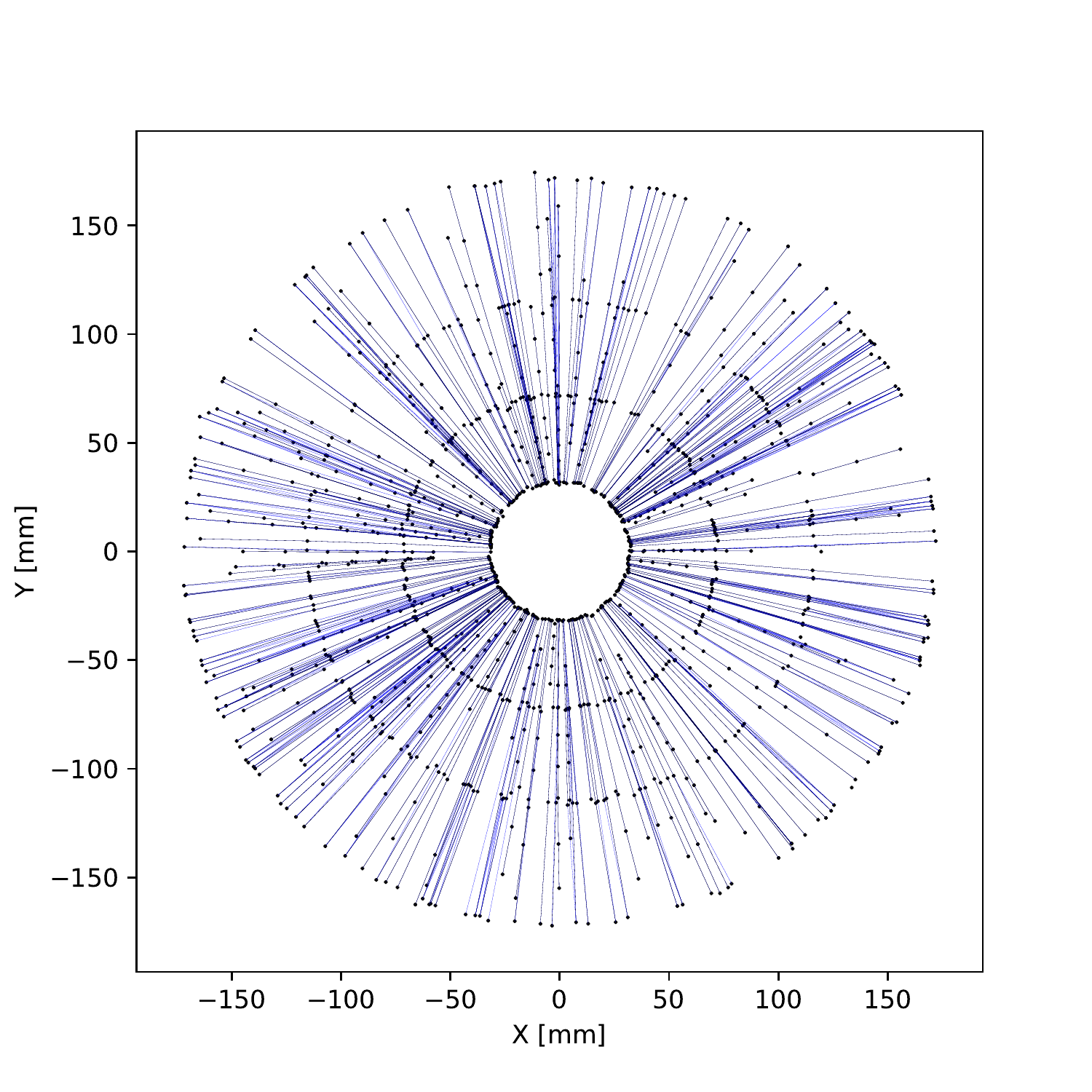}
    \includegraphics[width=0.33\textwidth,clip=true,viewport = 0 10 392 382]{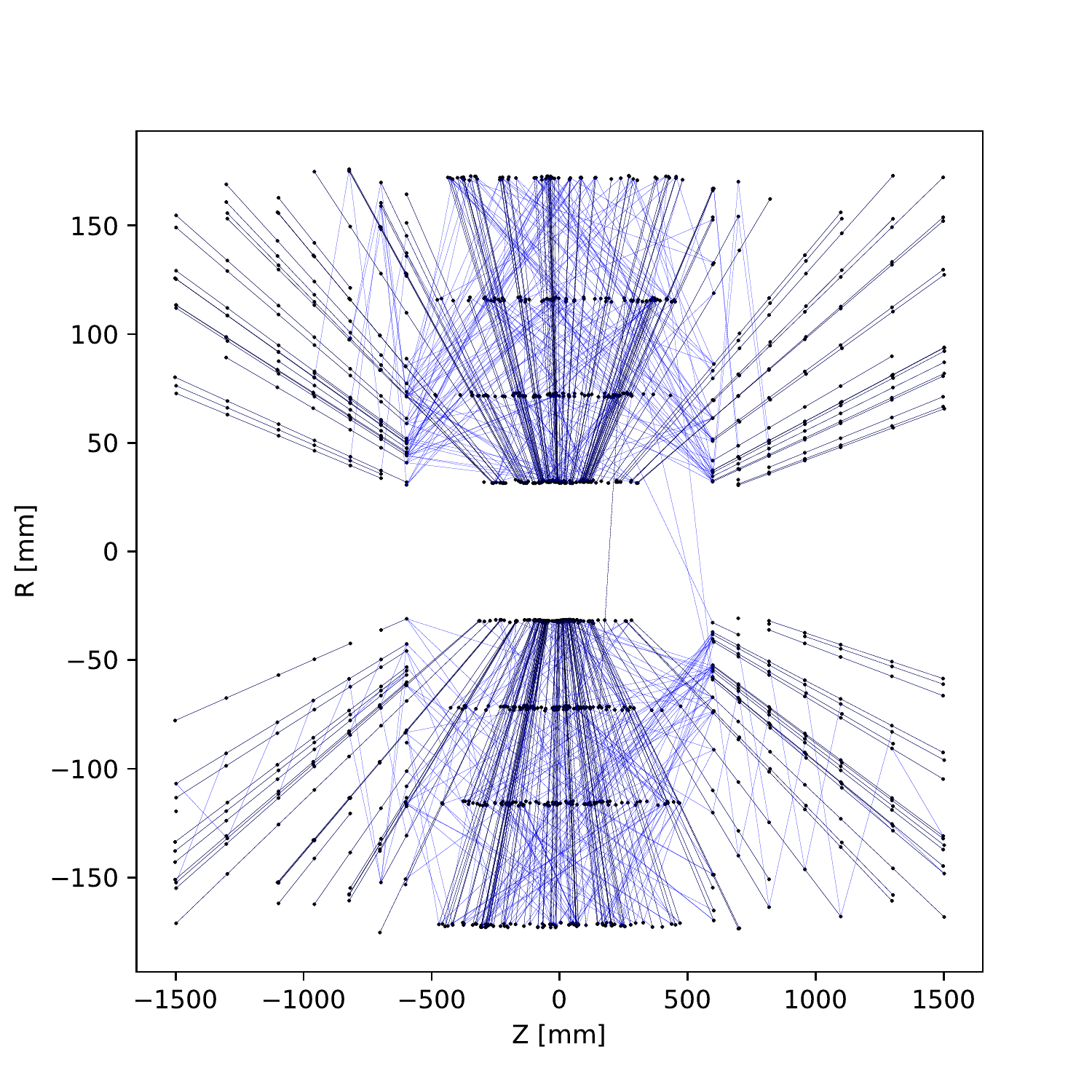}\\~\\
    \includegraphics[height=2.8cm,clip=true,viewport = 0 270 1920 710]{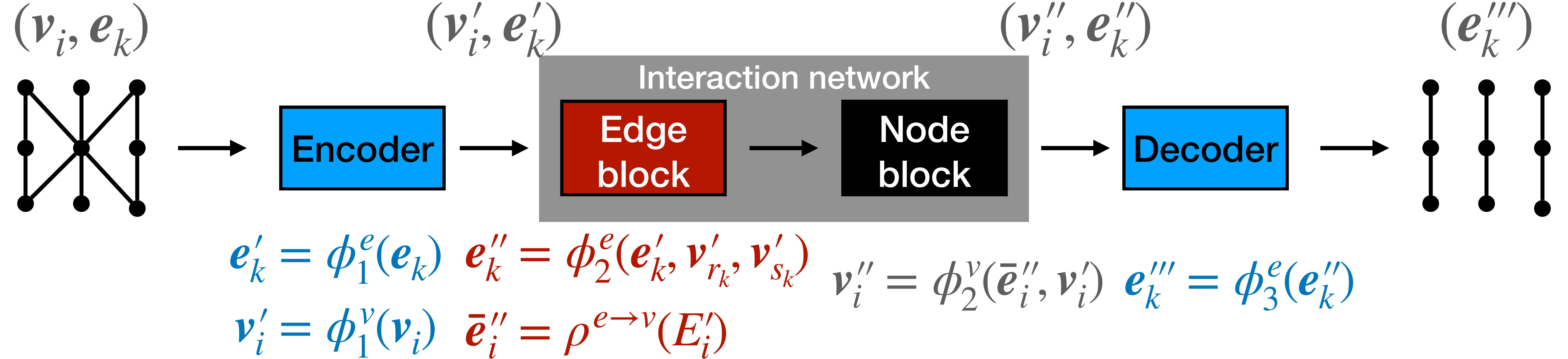}\\[3ex]
    \includegraphics[height=2.8cm,clip=true,viewport = 150 270 1770 710]{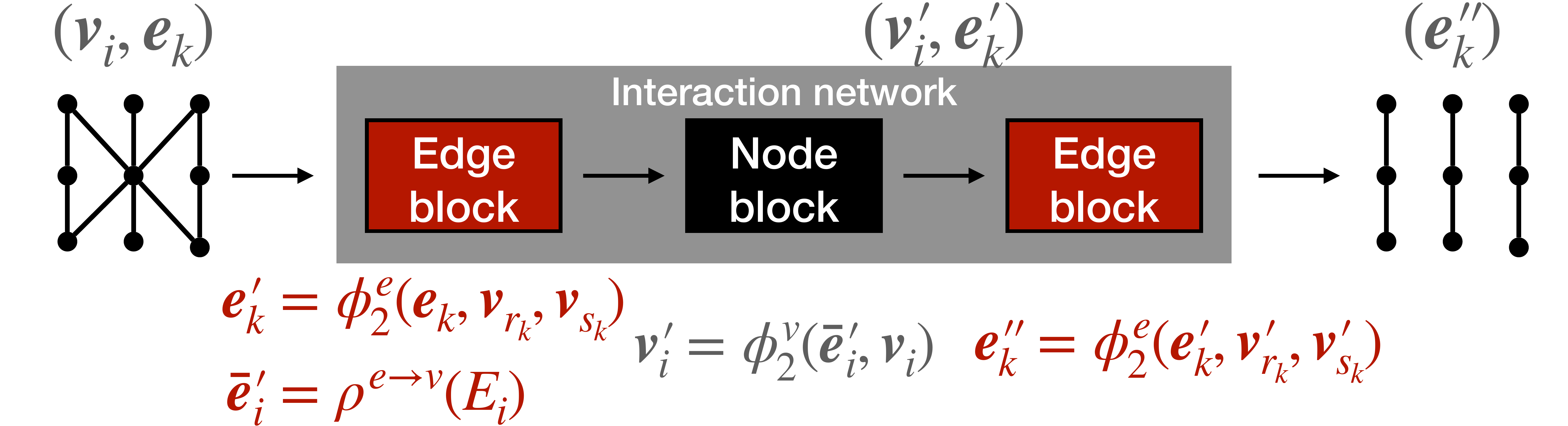}
    \caption{
    %TrackML detector regions labeled 7, 8, and 9, used in this study, consisting of 7, 4, and 4 layers, respectively (upper). 
    A characteristic graph of particles with $\pt>2$~GeV for one event in $x$--$y$ view (upper left) and $r$--$z$ view (upper right).
    The black edges correspond to true track segments, while the blue edges are spurious.
    %While the OpenCL implementation uses the full graphs, the \hlsfml implementation considers segments of the graph corresponding to 8 sectors in $\phi$, and 2 sectors in $\eta$.
    GNN architectures used for the \hlsfml (middle) and OpenCL (lower) implementations.}
    \label{fig:graphs}
\end{figure}

We focus the scope of the task by considering the innermost layers of the detector, which usually correspond to the ``pixel'' detector of a modern HEP detector. 
We consider two subdetector regions in the endcaps, consisting of 7 layers each, and one in the barrel of the TrackML detector, consisting of 4 layers.
Edges are constructed from hit pairs on adjacent layers satisfying $\Delta z < 15$~cm and $\Delta\phi / \Delta r < 2.62\times 10^{-4}$.
Particles with transverse momentum ($\pt$) greater than a given threshold are used to define the hits (nodes) and true edges.
For the \hlsfml model, we consider graphs corresponding to particle $\pt>2$~GeV, while for the OpenCL implementation we study the scaling as a function of $\pt$ from 1 to 5~GeV. 
A characteristic graph for a single event is shown in Fig.~\ref{fig:graphs} for $\pt>2$~GeV.
The graph size depends strongly on the minimum $\pt$.
For $\pt > 1$~GeV ($2$~GeV), the event graph contains approximately 5,300 (1,100) nodes and 16,600 (1,500) edges on average.
%For instance, just the neural network inference in the \hlsfml model for a smaller sectorized graph requires  226~kFLOP.
%\TODO{insert some numbers, average node/edges vs $\pt$}

Following the approach of Ref.~\cite{Farrell:2018cjr,Ju:2020xty}, we define a ``segment classifier'' using an IN model~\cite{Battaglia:2016jem,battaglia2018relational} to learn which edges connect hits belonging to the same track.
Relative to Ref.~\cite{Ju:2020xty}, the model architecture is simplified for the more limited task and the FPGA implementations. 
To implement it with \hlsfml, the model shown in Figure~\ref{fig:graphs} (lower left) is used.
The same type of model is used for the OpenCL implementation, as shown in Figure~\ref{fig:graphs} (lower right), without encoder and decoder networks and a different IN structure.
In addition, while both models consider the 3 input node features $(r, \phi, z)$, the \hlsfml model considers 4 input edge features $(\Delta r, \Delta\phi, \Delta z, \Delta R)$, while the OpenCL model does not consider any.
Instead, in this model, the input edge features are zero vectors $\boldsymbol{e}_k = \boldsymbol{0}$, which can be thought of as the ``initial guess'' of the edge weights, i.e. all edges are initially assumed to be fake.
Though we show results separately for these specific versions of the models, the \hlsfml and OpenCL implementations we developed are modular and flexible enough to accommodate both models as well as other permutations.
Appendix~\ref{sec:hls4mlmodelv2} demonstrates this by implementing a smaller version of the second model in \hlsfml.
Benchmarking the same model implemented in both frameworks is planned for future work.

For the \hlsfml model, an encoder composed of two neural networks, $\phi_1^e$ and $\phi_1^v$, transforms input node and edge features into hidden representations.
Both $\phi_1^e$ and $\phi_1^v$ have two layers with 8 neurons each and rectified linear unit (ReLU)~\cite{relu1,relu2} activation functions.
The IN is divided into two parts: an \emph{edge block} (or \emph{relational model}) and a \emph{node block} (or \emph{object model}).
The edge block network $\phi_2^e$ takes as input a pair of node features with the corresponding edge features and has layers of sizes $(8, 8)$ with ReLU activation.
The outputs of the $\phi_2^e$ are updated edge features, which can be considered ``messages'' sent between the nodes and aggregated at each node.
The node block network $\phi_2^v$ takes as input the aggregated messages and the updated node features and consists of layers of sizes $(8, 8)$ with ReLU activation.
Finally, the decoder network $\phi_3^v$ transforms edge features into an edge weight classifier (the probability for a given edge to be a true track segment), and has layers of sizes $(8, 8, 8, 1)$ with ReLU activation on all but the final layer that has a sigmoid activation.

For the model used in the OpenCL implementation, the IN begins with an edge block where $\phi_2^e$ has layers of sizes $(250, 250, 250, 1)$ with ReLU activations except for the final layer, which has a sigmoid activation.
This is followed by a node block where $\phi_2^v$ has layers of sizes $(200, 200, 3)$ with ReLU activations except for the final layer.
Subsequently, the same edge block is repeated to calculate the edge weights.
The number of floating point operations (FLOPs) for both of these models scales with the number of nodes, number of edges, and size of the neural networks. 

\section{Implementations}
\label{sec:impl}

OpenCL is an open-source C-based interface for parallel computing on various hardware platforms, including CPUs, GPUs, digital signal processors (DSPs), and FPGAs, using task and data-based parallelism~\cite{OpenCL}.
%It is an industry standard language to write hardware description language (HDL) at a high level, making it easier to program hardware accelerators. 
%As a non-proprietary language, OpenCL provides a flexible and common interface across many types of hardware accelerators, greatly improving its potential use in a broad range of contexts.
The OpenCL implementation of the IN adopts a CPU-plus-FPGA coprocessing approach where the host program on the CPU manages the application, and all computational operations are accelerated using dedicated kernels deployed on the FPGA that take capitalize on the device's hardware architecture to parallelize operations. 
%These kernels are optimized during compilation for efficient execution. 
%These kernels take advantage of the device's hardware architecture to parallelize task- and data-based operations. 
%In our implementation of the IN, each kernel is implemented as an NDRange kernel, which provides explicit implementation of data parallelism during computation. 
%% maybe too much detail:
%NDRange kernels rely on partitioning data among work-items on the device. 
%For example, each operation in a loop is executed on an individual work item in the work group.
The matrix multiplication kernel is repeatedly executed during a forward pass of the network and leverages the FPGA architecture for an efficient data-parallel implementation. 
This kernel uses both 2D local memory tiling and 2D register blocking to reduce the redundancy and latency of reading from globally shared off-chip memory. % that is inherent to a naive matrix multiplication example. 
%2D memory tiling loads data from off-chip globally shared memory to local memory and computation is, therefore, based on local memory data. 
%Register blocking goes further by sub-tiling the local memory and storing values into registers so that more work per thread is possible. 
%Therefore, computation requires less access to local memory, leading to a higher percentage of fused multiply-accumulate (FMA) instructions. 
Because the input graph sizes to the network changes per event, the matrix multiplication kernels pad each matrix before computing the result. 
Throughout the forward pass, several optimizations are made to speed up computation. 
%The design process on optimizing the program focused on improving matrix multiplication operations, lowering device resource usage, and leveraging the inherent efficiencies of OpenCL.
One example is the use of double buffering, which allows kernels to transfer data and execute instructions concurrently.
%data relevant to the next kernel to be transferred while a preceding kernel is under execution. 
%Therefore, runtimes are reduced by simultaneously executing kernels while transferring data in preparation of executing the next kernel.
All loops iterations executed in the OpenCL kernels are ``unrolled'' to run in parallel, which decreases the latency at the cost of increased hardware resource consumption.
%performed in parallel  unrolled to decrease the number of iterations that the offline compiler must execute. 
%Although this comes at the expense of increased hardware resource consumption, kernel performance increases by decreasing the number of iterations of a loop the kernel executes.
The OpenCL implementation is tested with %a 4-rack Intel Xeon computer cluster as the host and 
an Intel Programmable Accelerator Card featuring an Arria 10 GX 1150 FPGA.

%\section{\hlsfml Implementation}
%\label{sec:hls4ml}
\hlsfml~\cite{Duarte:2018ite} is an open-source converter of ML models into FPGA firmware utilizing Xilinx Vitis~\cite{vitis20192} high-level synthesis (HLS)~\cite{vivadohls} for both pure FPGA hardware applications and coprocessing kernels. 
%HLS was created to accelerate the process of programming FPGAs, and \hlsfml provides an extra layer of abstraction to support ML.
%\hlsfml reads in model architectures and weights trained with standard ML libraries and converts them to HLS. 
%Included with \hlsfml is a library of common neural network building blocks which aid in the model conversion. 
%From the HLS model, the vendor tools create a firmware design. 
In addition to implementing a GNN using available \hlsfml tools, efforts are in progress to extend the compiler to support basic GNN building blocks.
%The \hlsfml implementation accesses the inherent parallezation of FPGAs through preprocessor directives, or \emph{pragmas}, to lower the latency of model evaluation. 
Each forward pass of IN is pipelined such that matrix multiplications are performed in parallel. 
%Trade-offs between latency and resource utilization limit the realistic depth of parallelization possible. 
%As such, the 
Pipelining is performed at the level of the IN edge and node blocks %, and not in the dense neural network layers. 
and the amount of pipelining is tunable through the \emph{reuse factor} parameter which controls the initiation interval (II) of each block.
%Smaller graphs are less resource intensive, so the IN can be pipelined completely with an II of 1, while for larger graphs, less pipelining (larger II) reduces the FPGA resource consumption
In conjunction with block-level pipelining, all loops are fully unrolled to decrease latency.
All GNN model inputs are implemented with a streaming interface, which creates a FIFO that recycles its storage of array elements over each passage of a neural network block. 
%Another feature of FPGAs implemented in the \hlsfml GNN is the streaming option. 
%Arrays are by default implemented in RAM. 
%Designating an array with a streaming pragma implements it as a FIFO that recycles its storage of array elements over each passage of a neural network block. 
Streaming decreases resource utilization but may increase latency. 
%However, the trade-off yields a net decrease in latency due to the concurrent pipelining. 
%The \hlsfml implementation includes streaming for all model inputs as an option.
The input graph size is truncated to 112 nodes and 148 edges, which corresponds to the 95th percentile graph size for the $\eta$ and $\phi$ sectors, and uses zero-padding to make the inputs a uniform size.
To make scans of resources and timing more tractable, we further subdivide these graphs into quarters of up to 28 nodes and 37 edges just for the purposes of presentation.
%To make scans of resources and timing more tractable, we  consider only up to 57 edges just for the purposes of presentation.
Tests of the \hlsfml implementation target a Xilinx Kintex UltraScale (KU) 115 FPGA.

\section{Results}
\label{sec:results}

\begin{figure}[htpb]
    \centering
    \includegraphics[width=0.33\textwidth]{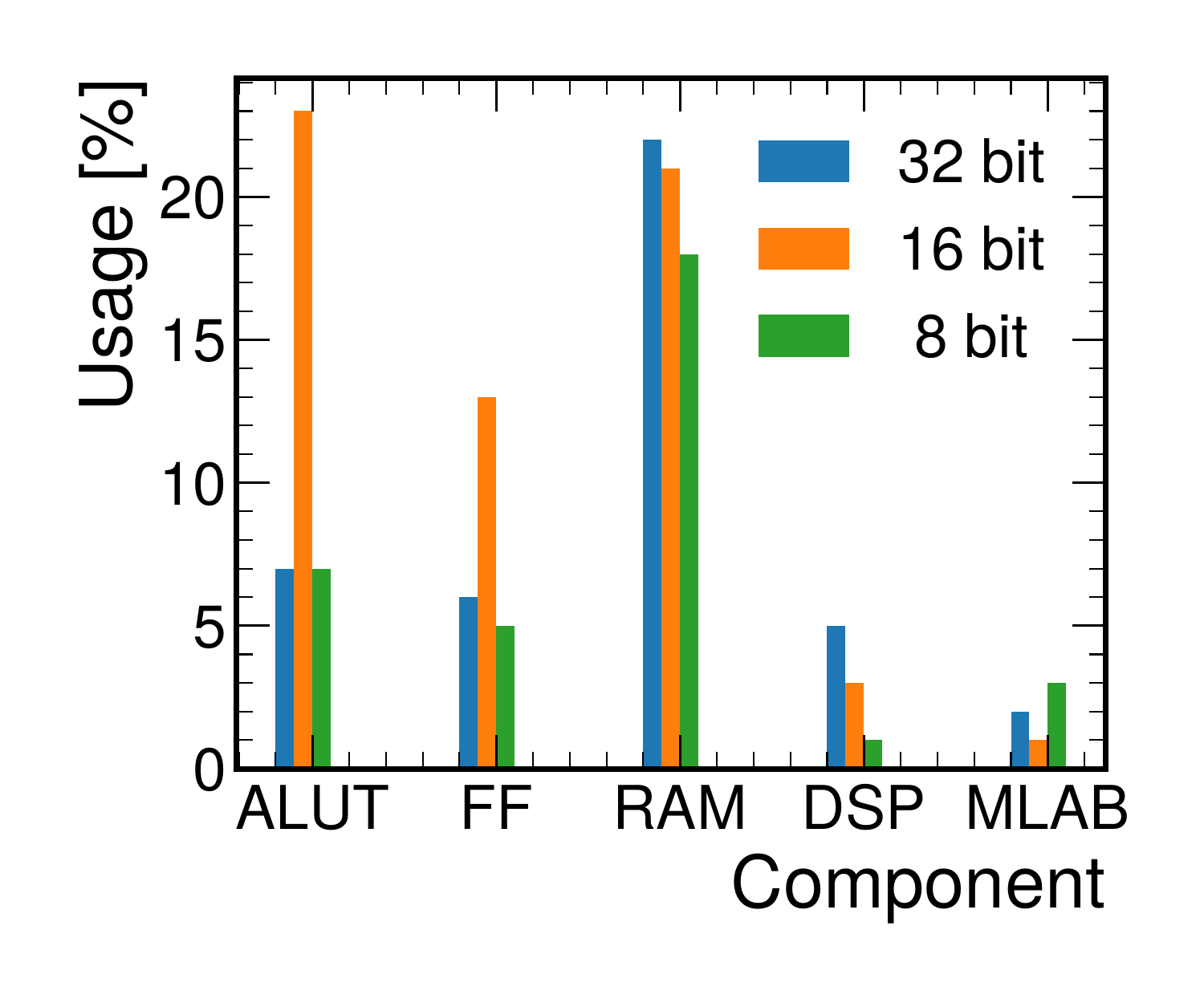}
    \includegraphics[width=0.66\textwidth]{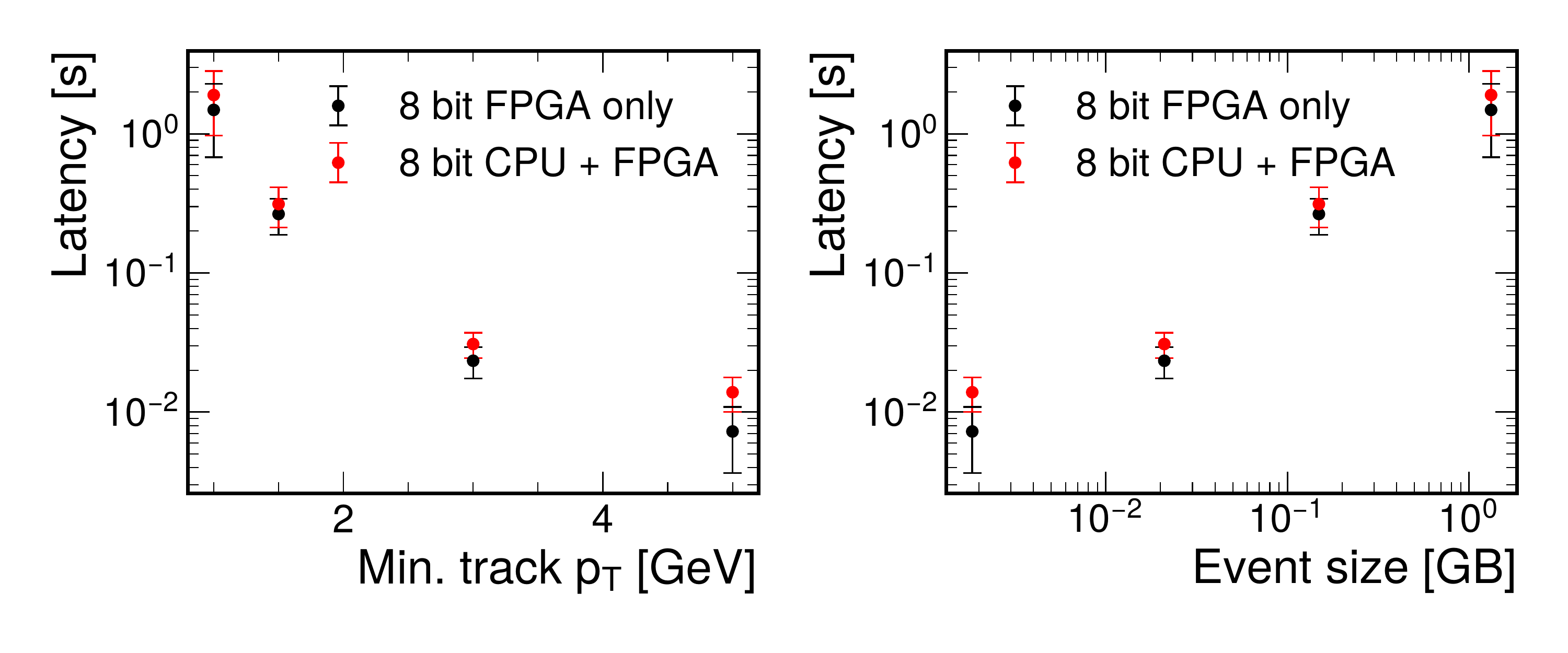}
    \caption{Resource system area analysis for the OpenCL implementation on an Arria 10 GX 1150 FPGA over multiple input data precision sizes (left). 
    Scalability study in terms of latency versus minimum $\pt$ (center) and event size (right).}
    \label{fig:ocl_precision_scalability}
\end{figure}

For the OpenCL implementation, Fig.~\ref{fig:ocl_precision_scalability} (left) compares the resource usage for input data represented with 8-, 16-, and 32-bit floating point precision.
For most components, operating with lower precision typically leads to lower resource usage, though the system area is dependent on other factors as well.
Execution times also increase with increasing precision from 8 to 32 bits, but not substantially, especially for lower minimum $\pt$.
%For larger minimum $\pt$ requirements, the bit precision does not show a significant difference, but as the minimum $\pt$ decreases, i.e. as the input data size increases, the difference gets larger. 
%The more significant metrics focus on the precision impacts on system area resource usage. 
%, the study demonstrates there are ways to implement the same algorithms with a lighter systematic load. 
Additionally, lower precision leads to smaller input data sizes, enabling the implementation to process events at lower $\pt$.

Figure~\ref{fig:ocl_precision_scalability} shows how the OpenCL implementation scales with minimum $\pt$ (center) and input data size (right).
Our implementation is able to process graphs with a minimum $\pt$ as low as 1~GeV, which is limited by hardware storage, such as host CPU RAM and on-device local RAM. 
An increase in input data size increases the execution times. 
The largest bottlenecks are matrix multiplication operations and the overhead time that accompanies the enqueueing the kernels. %enqueueing NDRange kernels.
Even though this overhead time increases as graph input size increases, enqueueing becomes more efficient. %, as Fig.~\ref{fig:ocl_precision_scalability} (center and right) demonstrates.
This scalability study shows that the coprocessing approach in OpenCL is flexible with respect to data size and can handle significant increases in data sizes.
However, we note that our CPU-FPGA coprocessing implementation is slower by about a factor of 10 from a pure CPU-based approach. 
In particular, CPU-based inference of the same model for $\pt > 1$~(5)~GeV graphs in \textsc{PyTorch}~\cite{pytorch} is about 86~(2)~ms.

\begin{figure}[htpb]
    \centering
    \includegraphics[width=0.24\textwidth]{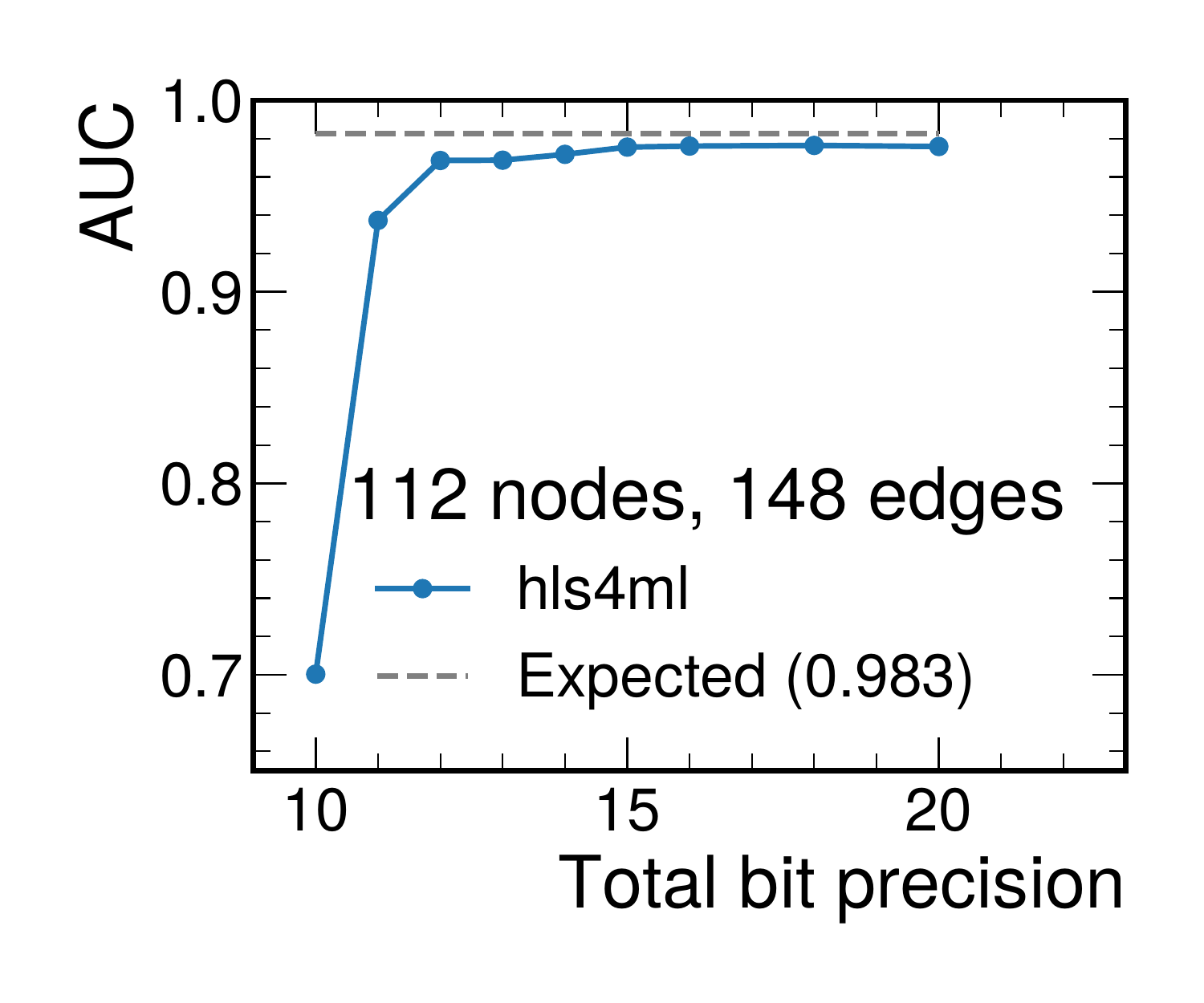}
    \includegraphics[width=0.24\textwidth]{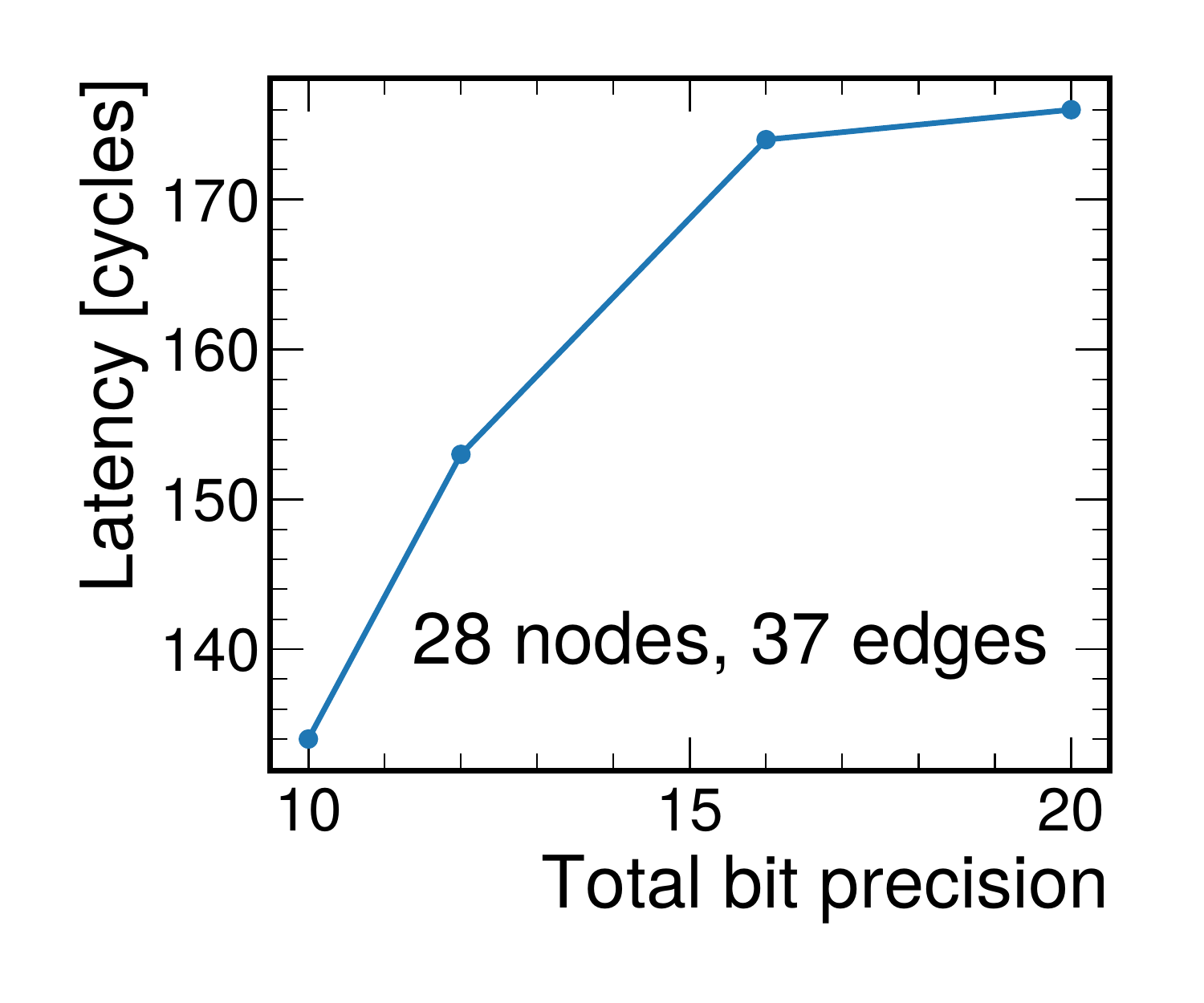}    %\includegraphics[width=0.24\textwidth]{figures/Resources_vs_BP.pdf}
    \includegraphics[width=0.24\textwidth]{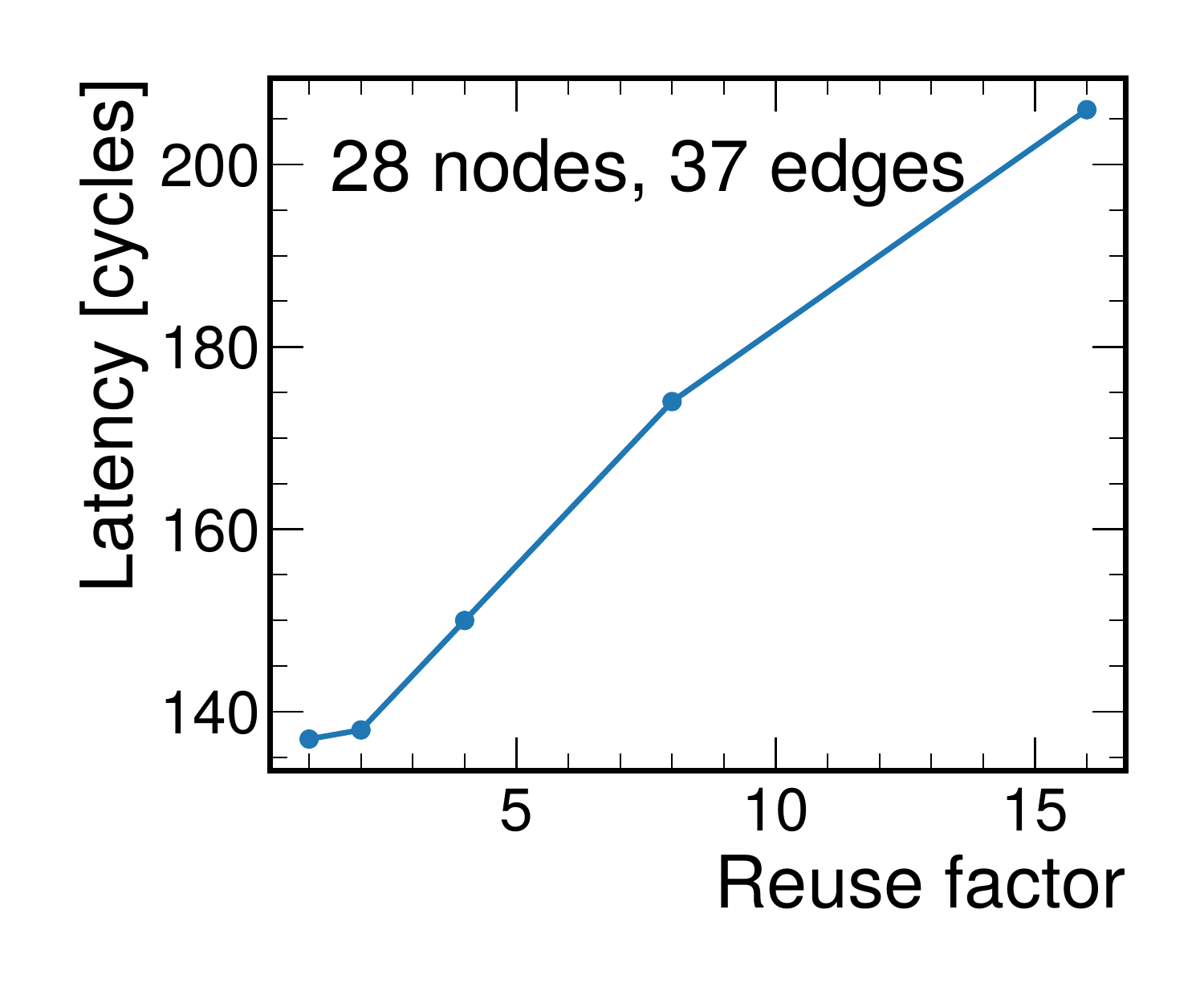}
    \includegraphics[width=0.24\textwidth]{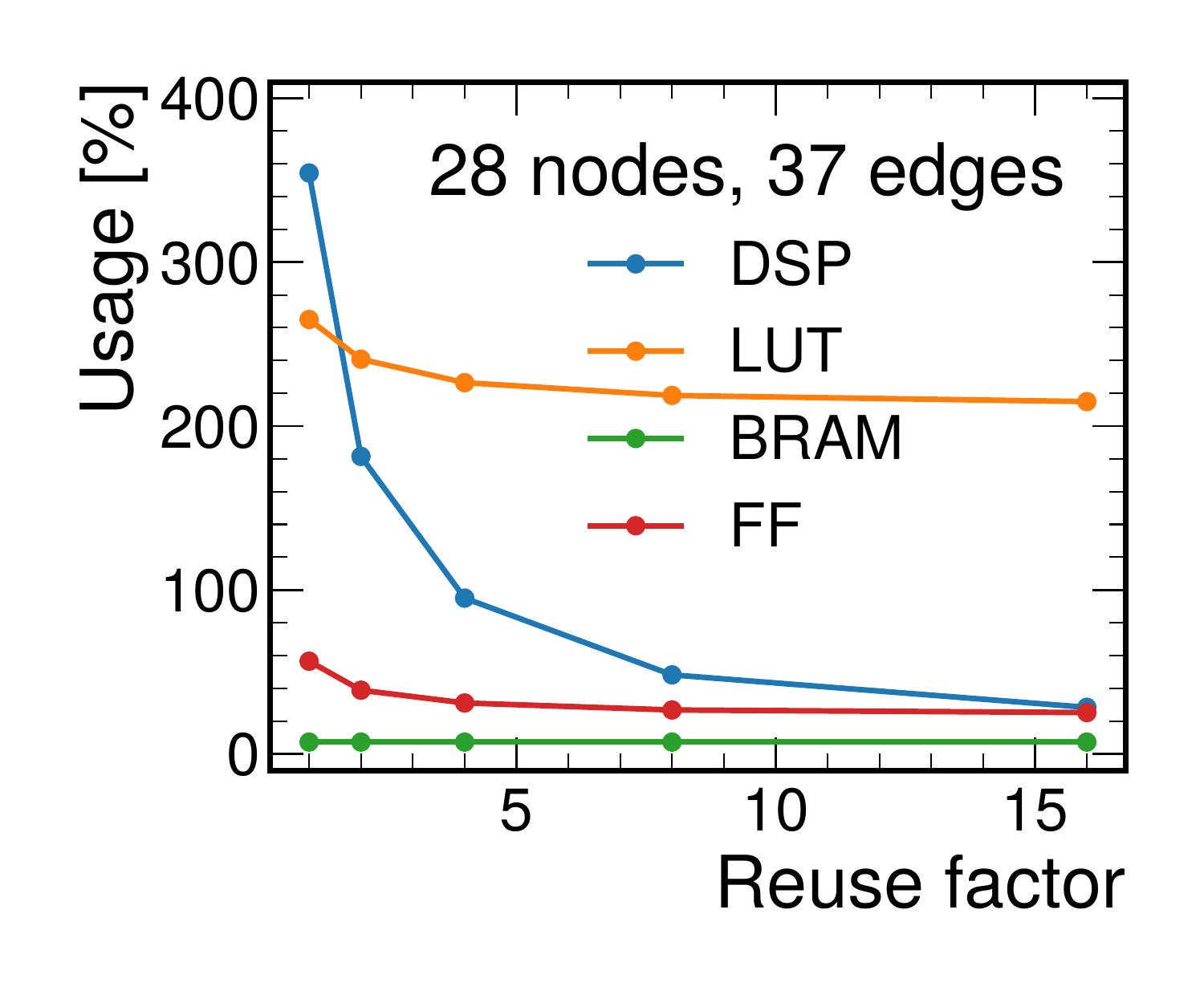}
    \caption{Segment classifier AUC versus fixed point precision total bit width (far left) for the \hlsfml implementation. 
    Latency in clock cycles for a 5~ns clock period (center left) as a function of the total bit width for the \hlsfml implementation. 
    The reuse factor (and thus II) is set to be 8.
    Latency (center right) and resource usage estimates (far right) relative to the available resources on a Xilinx KU 115 FPGA versus reuse factor for fixed total bit precision of $\langle 16, 6\rangle$.}
    \label{fig:hlsfml_scan}
\end{figure}

For the \hlsfml implementation, we first scan the fixed point precision total bit width to determine its impact on the physics performance of the algorithm as well as the latency and resource usage on the FPGA. 
We evaluate the receiver operating characteristic (ROC) curve for the segment classifier, and use the area under the curve (AUC) as a performance metric. 
Figure~\ref{fig:hlsfml_scan} (far left) shows the AUC as a function of the total bit precision, while the integer part is fixed to 6 bits.
We see that with 12 total bits, we effectively reproduce the 32-bit floating point model.
Figure~\ref{fig:hlsfml_scan} (center left) also shows the latency in clock cycles (for a 5~ns clock period) as a function of the total bit precision, which ranges from about 650~ns to 1~$\mu$s.
For CPU-based inference of the same model, the latency is about 27~ms for the same sectorized $\pt > 2$~GeV graph in the \texttt{graph\_nets} framework~\cite{graph_nets} based on \textsc{TensorFlow}~\cite{tensorflow2015-whitepaper}.
We also scan the reuse factor at constant fixed point precision of $\langle16,6\rangle$, to study the resources and timing as a function of decreasing concurrency.
Figure~\ref{fig:hlsfml_scan} shows the latency (center left) and resource usage estimates (far right) versus reuse factor.
By construction, the II for the algorithm is equal to the reuse factor.
We note that the HLS lookup table (LUT) usage tends to be overestimated, while the HLS DSP usage tends to be accurate~\cite{Duarte:2018ite,Iiyama:2020wap,DiGuglielmo:2020eqx}.
Nonetheless, to make the algorithm fit on a single FPGA, larger reuse factors or bigger FPGAs may be necessary.

\section{Summary and Outlook}
\label{sec:summary}

We develop and study two complementary FPGA implementations of algorithms for charged particle tracking based on graph neural networks. 
The first, using OpenCL, targets CPU-FPGA coprocessing applications and achieves a latency of between $10$~ms--$1$~s depending on the minimum $\pt$ for full event graphs, including data transfer and I/O, for the model under consideration.
The second, using \hlsfml, targets both coprocessing and custom trigger (ultra low latency) applications and has an expected latency of 650~ns--1~$\mu$s, considering only the execution time on the FPGA, for smaller, sectorized input graphs and a more compact model, although more work is needed to reduce the resource consumption.
Compared to CPU-based execution of the same models, the speedup for the \hlsfml implementation is considerable, but further optimizations are needed to improve on the CPU latency for the OpenCL implementation.
Continued development in this direction may allow such algorithms to be used effectively in future computing workflows~\cite{Rankin:2020usv} and the Level-1 trigger at the LHC.
In future work, we plan to study detailed comparisons of the two implementations based on the same model, as well as comparing to GPU coprocessors~\cite{Krupa:2020bwg}.
Other optimizations of the GNN model may also be possible, such as more efficient architectures~\cite{Iiyama:2020wap} and use of quantization-aware training~\cite{DiGuglielmo:2020eqx,Coelho:2020zfu} to reduce the necessary precision.

\section*{Broader Impact}

This work may be used to accelerate particle tracking and other reconstruction algorithms in high energy physics experiments.
While accelerated machine learning on FPGAs has many potential benefits to science and society, including high-quality, fast data reconstruction and selection in experiments, automated detector control, or smarter IoT devices, it may also be used for nefarious purposes, such as surveillance or military unmanned aerial vehicles.
%Authors are required to include a statement of the broader impact of their work, including its ethical aspects and future societal consequences. 
%Authors should discuss both positive and negative outcomes, if any. For instance, authors should discuss a) who may benefit from this research, b) who may be put at disadvantage from this research, c) what are the consequences of failure of the system, and d) whether the task/method leverages biases in the data. 
%If authors believe this is not applicable to them, authors can simply state this.

%Use unnumbered first level headings for this section, which should go at the end of the paper. 
%{\bf Note that this section does not count towards the eight pages of content that are allowed.}

\begin{ack}

%% Fast ML acknowledgement:
We acknowledge the Fast Machine Learning collective as an open community of multi-domain experts and collaborators. 
This community was important for the development of this project. 
%% PICSciE acknowledgement:
The simulations presented for the OpenCL implementation in this article were performed on computational resources managed and supported by Princeton Research Computing, a consortium of groups including the Princeton Institute for Computational Science and Engineering (PICSciE) and the Office of Information Technology's High Performance Computing Center and Visualization Laboratory at Princeton University.
%% PRP Nautilus acknowledgement:
Work for the \texttt{hls4ml} implementation was partially performed on the Pacific Research Platform Nautilus HyperCluster supported by NSF awards CNS-1730158, ACI-1540112, ACI-1541349, OAC-1826967, the University of California Office of the President, and the University of California San Diego's California Institute for Telecommunications and Information Technology/Qualcomm Institute. 
Thanks to CENIC for the 100~Gpbs networks.

%% IRIS-HEP acknowledgement:
A.~H., V.~R., and S.~T. are supported by IRIS-HEP through the U.S. National Science Foundation (NSF) under Cooperative Agreement OAC-1836650.
%% Others:
J.~D. is supported by the U.S. Department of Energy (DOE), Office of Science, Office of High Energy Physics Early Career Research program under Award No. DE-SC0021187.
T.~A., V.~L., M.~P., and S.~S. are supported by the European Research Council (ERC) under the European Union's Horizon 2020 research and innovation program (Grant Agreement No. 772369).
L.~G., S.~J., and N.~T. are supported by Fermi Research Alliance, LLC under Contract No. DE-AC02-07CH11359 with the U.S. Department of Energy (DOE), Office of Science, Office of High Energy Physics.
P.~H. is supported by a Massachusetts Institute of Technology University grant. 
Z.~W. is supported by the National Science Foundation under Grants No. 1606321 and 115164.

\end{ack}

\appendix

\section{Alternative GNN Model in \hlsfml}
\label{sec:hls4mlmodelv2}

\begin{figure}[htpb]
    \centering
    \includegraphics[width=0.24\textwidth]{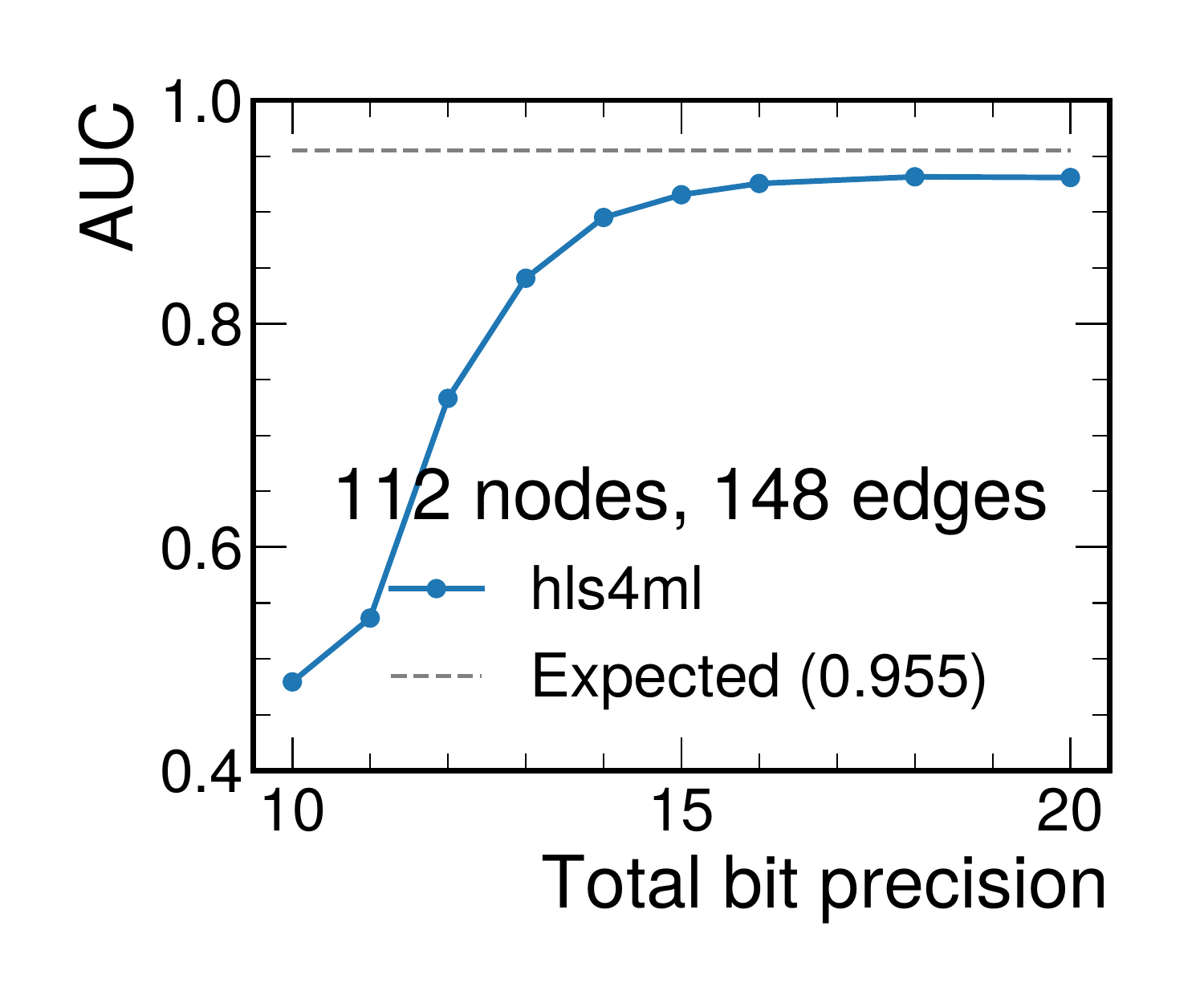}
    \includegraphics[width=0.24\textwidth]{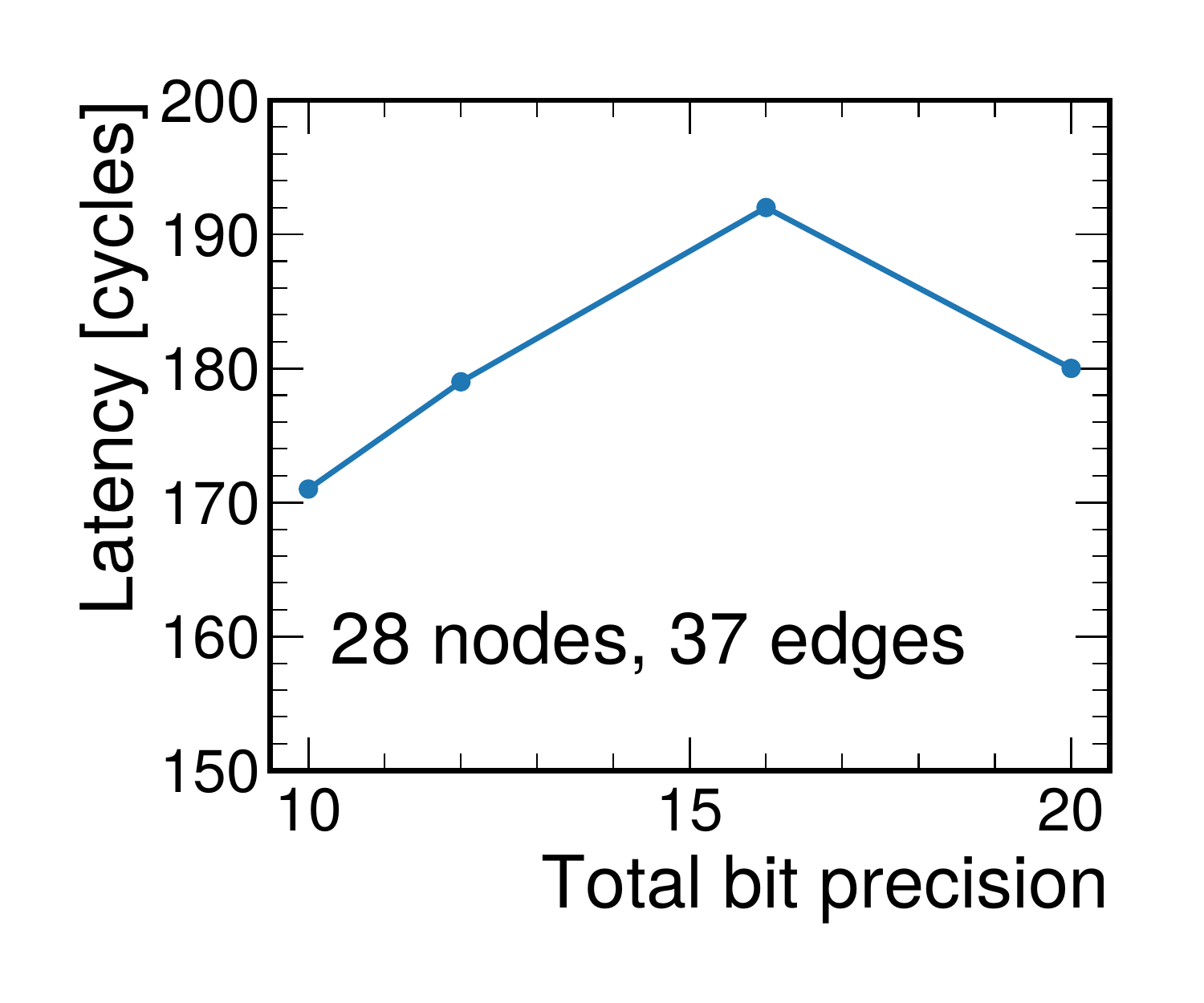}    %\includegraphics[width=0.24\textwidth]{figures/Resources_vs_BP_v2.pdf}
    \includegraphics[width=0.24\textwidth]{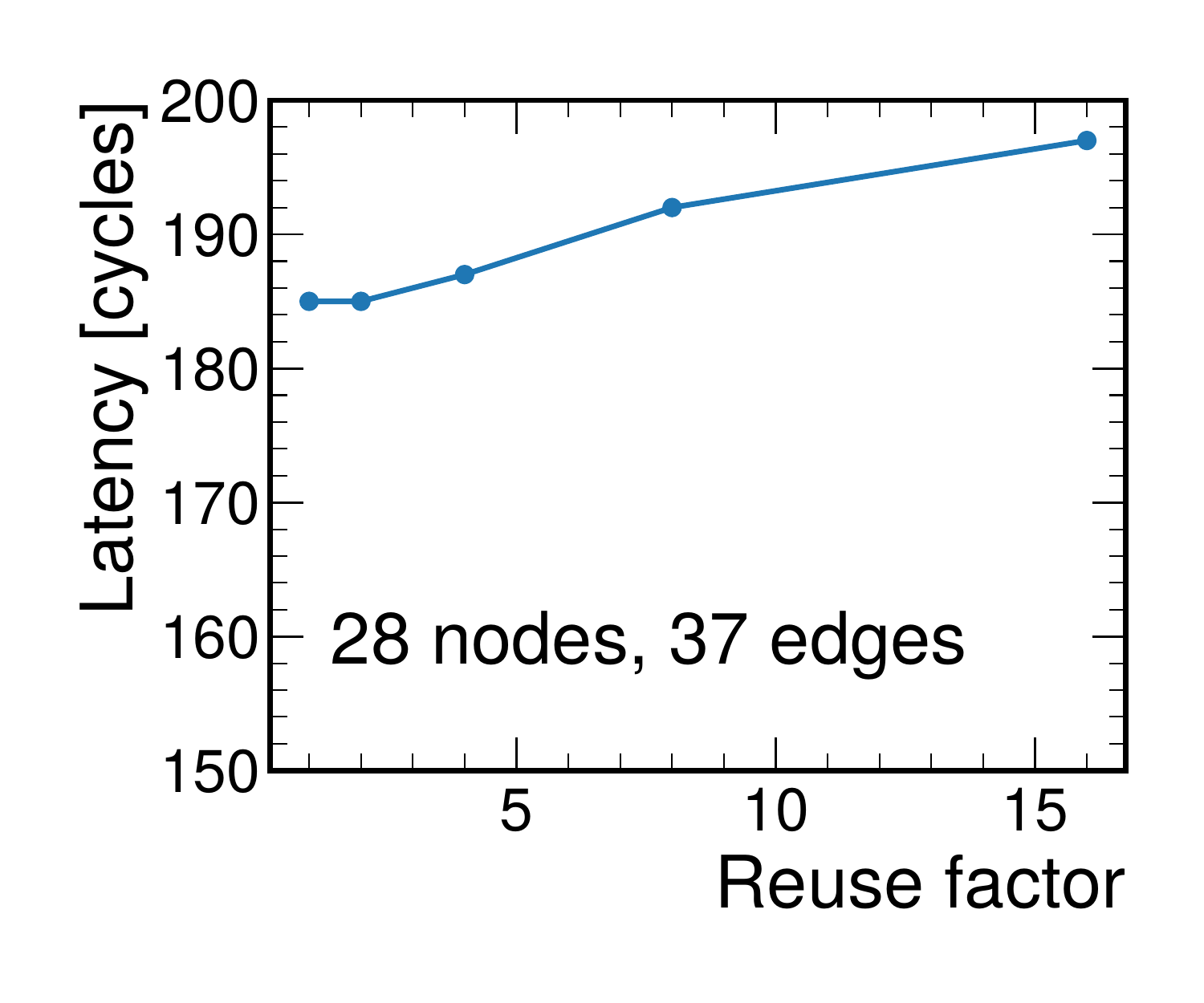}
    \includegraphics[width=0.24\textwidth]{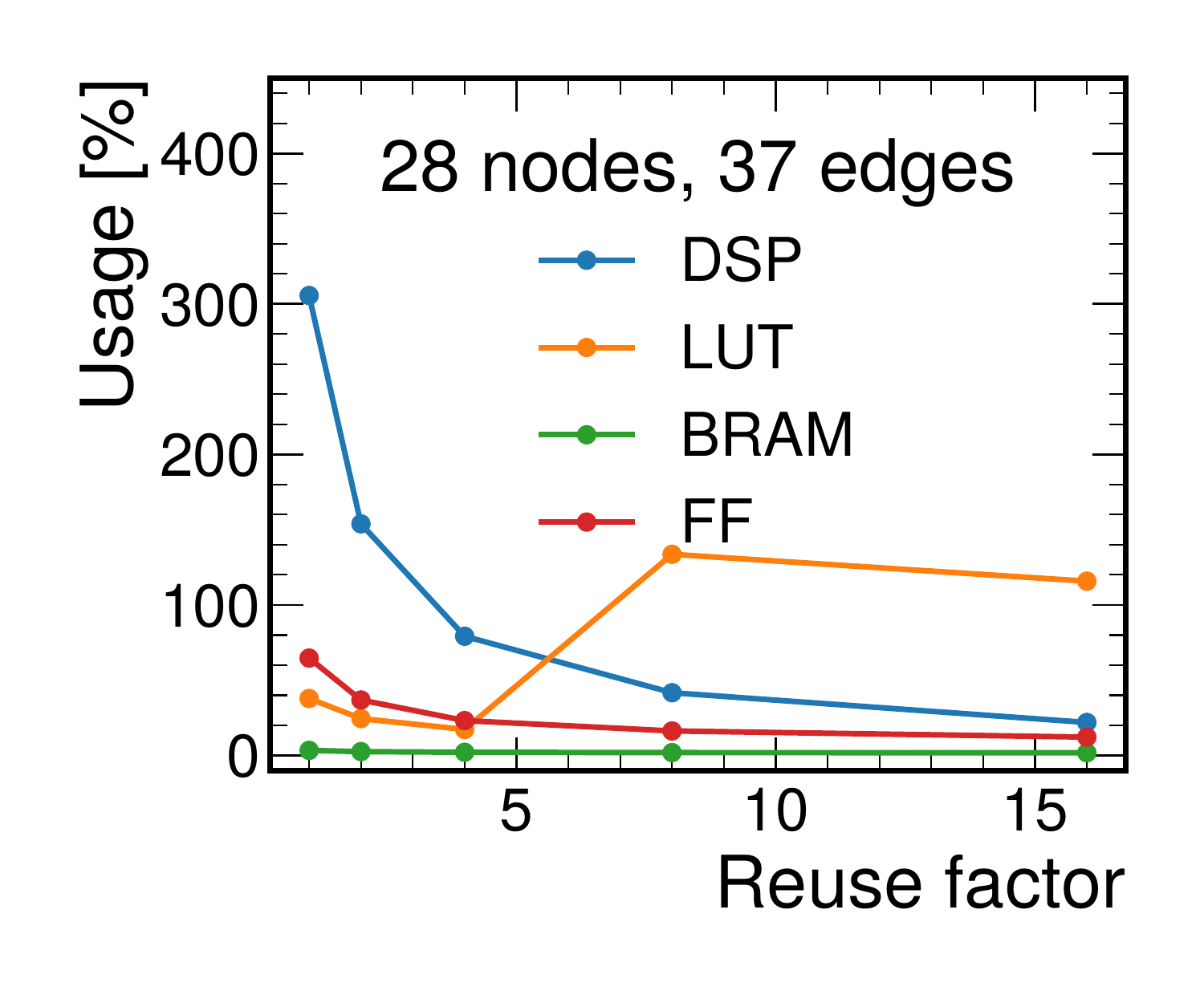}
    \caption{Alternative GNN segment classifier model implemented in \hlsfml.
    AUC (far left) and latency (center left) in clock cycles for a 5~ns clock period as a function of the total bit width.
    The reuse factor (and thus II) is set to be 8.
    Latency (center right) and resource usage estimates (far right) relative to the available resources on a Xilinx KU 115 FPGA versus reuse factor for fixed total bit precision of $\langle 16, 6\rangle$.}
    \label{fig:hlsfml_scan_v2}
\end{figure}

In addition to the first model shown in Fig.~\ref{fig:graphs}, we also implement the second model in \hlsfml.
In this model, the same architecture is used as the one implemented in OpenCL, except the neural network sizes are reduced: the $\phi_2^e$ network has layers of sizes $(8, 8, 8, 1)$ and $\phi_2^v$ has layers of sizes $(8, 8, 3)$.
Figure~\ref{fig:hlsfml_scan_v2} shows the AUC (far left) and the latency in clock cycles for a 5~ns clock period (center left) as a function of the total bit precision, while the integer part is fixed to 6 bits.
We see that above 16 total bits, the quantized model effectively reproduces the 32-bit floating point model.
Figure~\ref{fig:hlsfml_scan_v2} also shows the latency (center left) and resource usage estimates (far right) versus reuse factor at a constant fixed point precision of $\langle16,6\rangle$.
This small implemented model reasonably fits on a single FPGA with a latency less than 1~$\mu$s.

\bibliographystyle{lucas_unsrt}
\bibliography{references}

\providecommand{\href}[2]{#2}\begingroup\raggedright\begin{thebibliography}{10}%
\makeatletter
\providecommand{\hrefCMSnoop }[0]{\@secondoftwo}%
\makeatother
\providecommand{\doi}{\texttt{doi:}\begingroup \urlstyle{tt}\Url}

\bibitem{Amrouche:2019wmx}
S.~Amrouche\hrefCMSnoop {}{ {et~al.}, ``The tracking machine learning
  challenge: Accuracy phase'',} in \textit{ The NeurIPS '18 Competition},
  S.~Escalera and R.~Herbrich, eds., p.~231.
\newblock Springer, Cham, Switzerland, 2020.
\newblock
  \href{http://www.arXiv.org/abs/1904.06778}{\texttt{arXiv:1904.06778}}.
\newblock
  \href{http://dx.doi.org/10.1007/978-3-030-29135-8_9}{\doi{10.1007/978-3-030-29135-8_9}}.

\bibitem{Strandlie:2010zz}
\hrefCMSnoop {}{A.~Strandlie and R.~Fr{\"{u}}hwirth, ``{Track and vertex
  reconstruction: From classical to adaptive methods}'',} \textit{ Rev. Mod.
  Phys.} \textbf{ 82} (2010) 1419,
  \href{http://dx.doi.org/10.1103/RevModPhys.82.1419}{\doi{10.1103/RevModPhys.82.1419}}.

\bibitem{Chatrchyan:2014fea}
\hrefCMSnoop {}{{CMS} Collaboration, ``{Description and performance of track
  and primary-vertex reconstruction with the CMS tracker}'',} \textit{ J.
  Instrum.} (2014) P10009,
  \href{http://dx.doi.org/10.1088/1748-0221/9/10/P10009}{\doi{10.1088/1748-0221/9/10/P10009}},
  \href{http://www.arXiv.org/abs/1405.6569}{\texttt{arXiv:1405.6569}}.

\bibitem{Aaboud:2017all}
\hrefCMSnoop {}{{ATLAS} Collaboration, ``Performance of the {ATLAS} track
  reconstruction algorithms in dense environments in {LHC Run} 2'',} \textit{
  Eur. Phys. J. C} \textbf{ 77} (2017) 673,
  \href{http://dx.doi.org/10.1140/epjc/s10052-017-5225-7}{\doi{10.1140/epjc/s10052-017-5225-7}},
  \href{http://www.arXiv.org/abs/1704.07983}{\texttt{arXiv:1704.07983}}.

\bibitem{combkalman1}
\hrefCMSnoop {}{P.~Billoir, ``Progressive track recognition with a
  {Kalman}-like fitting procedure'',} \textit{ Comput. Phys. Comm.} \textbf{
  57} (1989) 390,
  \href{http://dx.doi.org/10.1016/0010-4655(89)90249-X}{\doi{10.1016/0010-4655(89)90249-X}}.

\bibitem{combkalman2}
\hrefCMSnoop {}{P.~Billoir and S.~Qian, ``Simultaneous pattern recognition and
  track fitting by the {Kalman} filtering method'',} \textit{ Nucl. Instrum.
  Methods Phys. Res. A} \textbf{ 294} (1990) 219,
  \href{http://dx.doi.org/10.1016/0168-9002(90)91835-Y}{\doi{10.1016/0168-9002(90)91835-Y}}.

\bibitem{combkalman3}
\hrefCMSnoop {}{R.~Mankel, ``A concurrent track evolution algorithm for pattern
  recognition in the hera-b main tracking system'',} \textit{ Nucl. Instrum.
  Methods Phys. Res. A} \textbf{ 395} (1997) 169,
  \href{http://dx.doi.org/10.1016/S0168-9002(97)00705-5}{\doi{10.1016/S0168-9002(97)00705-5}}.

\bibitem{kalman}
\hrefCMSnoop {}{{R. Fr\"{u}hwirth}, ``Application of {Kalman} filtering to
  track and vertex fitting'',} \textit{ Nucl. Instrum. Methods Phys. Res. A}
  \textbf{ 262} (1987) 444,
  \href{http://dx.doi.org/10.1016/0168-9002(87)90887-4}{\doi{10.1016/0168-9002(87)90887-4}}.

\bibitem{breakdown}
H.~Esmaeilzadeh\hrefCMSnoop {}{ {et~al.}, ``Dark silicon and the end of
  multicore scaling'',} in \textit{ Proceedings of the 38th Annual
  International Symposium on Computer Architecture}, p.~365.
\newblock ACM, New York, NY, USA, 2011.
\newblock
  \href{http://dx.doi.org/10.1145/2000064.2000108}{\doi{10.1145/2000064.2000108}}.

\bibitem{dennard}
R.~H. {Dennard}\hrefCMSnoop {}{ {et~al.}, ``Design of ion-implanted {MOSFET}'s
  with very small physical dimensions'',} \textit{ IEEE J. Solid-State
  Circuits} \textbf{ 9} (1974) 256,
  \href{http://dx.doi.org/10.1109/JSSC.1974.1050511}{\doi{10.1109/JSSC.1974.1050511}}.

\bibitem{Farrell:2018cjr}
\hrefCMSnoop {}{S.~Farrell {et~al.}, ``{Novel deep learning methods for track
  reconstruction}'',} in \textit{ {4th International Workshop Connecting The
  Dots 2018}}.
\newblock 2018.
\newblock
\href{http://www.arXiv.org/abs/1810.06111}{\texttt{arXiv:1810.06111}}.
\newblock
%%CITATION = ARXIV:1810.06111;%%.

\bibitem{Ju:2020xty}
\href
  {https://ml4physicalsciences.github.io/files/NeurIPS_ML4PS_2019_83.pdf}{X.~Ju
  {et~al.}, ``Graph neural networks for particle reconstruction in high energy
  physics detectors'',} in \textit{ {Machine Learning and the Physical Sciences
  Workshop at the 33rd Annual Conference on Neural Information Processing
  Systems}}.
\newblock 2019.
\newblock
  \href{http://www.arXiv.org/abs/2003.11603}{\texttt{arXiv:2003.11603}}.

\bibitem{Shlomi:2020gdn}
\hrefCMSnoop {}{J.~Shlomi, P.~Battaglia, and J.-R. Vlimant, ``Graph neural
  networks in particle physics'',}
  \href{http://dx.doi.org/10.1088/2632-2153/abbf9a}{\doi{10.1088/2632-2153/abbf9a}},
  \href{http://www.arXiv.org/abs/2007.13681}{\texttt{arXiv:2007.13681}}.
  Accepted by \emph{Mach. Learn.: Sci. Tech.}

\bibitem{Battaglia:2016jem}
P.~W. Battaglia\hrefCMSnoop {}{ {et~al.}, ``Interaction networks for learning
  about objects, relations and physics'',} in \textit{ {Advances in Neural
  Information Processing Systems}}, volume~29, p.~4502.
\newblock 2016.
\newblock
  \href{http://www.arXiv.org/abs/1612.00222}{\texttt{arXiv:1612.00222}}.

\bibitem{battaglia2018relational}
\hrefCMSnoop {}{P.~W. Battaglia {et~al.}, ``Relational inductive biases, deep
  learning, and graph networks'',}
  \href{http://www.arXiv.org/abs/1806.01261}{\texttt{arXiv:1806.01261}}.

\bibitem{ATLASL1T}
\hrefCMSnoop {}{{ATLAS} Collaboration, ``{Operation of the ATLAS trigger system
  in Run 2}'',} \textit{ J. Instrum.} \textbf{ 15} (2020) P10004,
  \href{http://dx.doi.org/10.1088/1748-0221/15/10/P10004}{\doi{10.1088/1748-0221/15/10/P10004}},
  \href{http://www.arXiv.org/abs/2007.12539}{\texttt{arXiv:2007.12539}}.

\bibitem{ATLASP2L1T}
\href {https://cds.cern.ch/record/2285584}{{ATLAS} Collaboration, ``{Technical
  Design Report for the Phase-II Upgrade of the ATLAS TDAQ System}'',} ATLAS
  Technical Design Report CERN-LHCC-2017-020. ATLAS-TDR-029, 2017.

\bibitem{CMSL1T}
\hrefCMSnoop {}{{CMS} Collaboration, ``{Performance of the CMS Level-1 trigger
  in proton-proton collisions at $\sqrt{s} =$ 13 TeV}'',} \textit{ J. Instrum.}
  \textbf{ 15} (2020) P10017,
  \href{http://dx.doi.org/10.1088/1748-0221/15/10/P10017}{\doi{10.1088/1748-0221/15/10/P10017}},
  \href{http://www.arXiv.org/abs/2006.10165}{\texttt{arXiv:2006.10165}}.

\bibitem{CMSP2L1T}
\href {https://cds.cern.ch/record/2714892}{{CMS} Collaboration, ``The {Phase-2}
  upgrade of the {CMS} {Level-1} trigger'',} CMS Technical Design Report
  CERN-LHCC-2020-004. CMS-TDR-021, 2020.

\bibitem{OpenCL}
\hrefCMSnoop {}{J.~E. {Stone}, D.~{Gohara}, and G.~{Shi}, ``{OpenCL}: A
  parallel programming standard for heterogeneous computing systems'',}
  \textit{ Comput. Sci. Eng.} \textbf{ 12} (2010) 66,
  \href{http://dx.doi.org/10.1109/MCSE.2010.69}{\doi{10.1109/MCSE.2010.69}}.

\bibitem{Duarte:2018ite}
\hrefCMSnoop {}{J.~Duarte {et~al.}, ``{Fast inference of deep neural networks
  in FPGAs for particle physics}'',} \textit{ J. Instrum.} \textbf{ 13} (2018)
  P07027,
  \href{http://dx.doi.org/10.1088/1748-0221/13/07/P07027}{\doi{10.1088/1748-0221/13/07/P07027}},
\href{http://www.arXiv.org/abs/1804.06913}{\texttt{arXiv:1804.06913}}.
%%CITATION = ARXIV:1804.06913;%%.

\bibitem{vloncar_2020_4161550}
V.~Loncar\hrefCMSnoop {}{ {et~al.}, ``fastmachinelearning/\texttt{hls4ml}:
  aster'',} 10, 2020.
\newblock
  \href{http://dx.doi.org/10.5281/zenodo.4161550}{\doi{10.5281/zenodo.4161550}},
  \url {https://github.com/fastmachinelearning/hls4ml}.

\bibitem{relu1}
\href {https://icml.cc/Conferences/2010/papers/432.pdf}{V.~Nair and G.~E.
  Hinton, ``Rectified linear units improve restricted {Boltzmann} machines'',}
  in \textit{ Proceedings of the 27th International Conference on International
  Conference on Machine Learning}, ICML'10, p.~807.
\newblock Omnipress, Madison, WI, USA, 2010.

\bibitem{relu2}
\href {http://proceedings.mlr.press/v15/glorot11a.html}{X.~Glorot, A.~Bordes,
  and Y.~Bengio, ``Deep sparse rectifier neural networks'',} in \textit{
  Proceedings of the 14th International Conference on Artificial Intelligence
  and Statistics}, G.~Gordon, D.~Dunson, and M.~Dud\'{i}k, eds., volume~15 of
  \textit{ Proceedings of Machine Learning Research}, p.~315.
\newblock JMLR, Fort Lauderdale, FL, USA, 4, 2011.

\bibitem{vitis20192}
\href
  {https://www.xilinx.com/support/documentation/sw_manuals/xilinx2019_2/ug902-vivado-high-level-synthesis.pdf}{{Xilinx,
  Inc.}, ``Vivado design suite user guide: High level synthesis'',} 2020.
\newblock \url
  {https://www.xilinx.com/support/documentation/sw_manuals/xilinx2019_2/ug902-vivado-high-level-synthesis.pdf}.

\bibitem{vivadohls}
D.~O'Loughlin\hrefCMSnoop {}{ {et~al.}, ``Xilinx vivado high level synthesis:
  Case studies'',} in \textit{ Irish Signals \& Systems Conference 2014 and
  2014 China-Ireland International Conference on Information and Communications
  Technologies (ISSC 2014/CIICT 2014). 25th IET Year}, p.~352.
\newblock 2014.
\newblock
  \href{http://dx.doi.org/10.1049/cp.2014.0713}{\doi{10.1049/cp.2014.0713}}.

\bibitem{pytorch}
A.~Paszke\href
  {http://papers.neurips.cc/paper/9015-pytorch-an-imperative-style-high-performance-deep-learning-library.pdf}{
  {et~al.}, ``{PyTorch}: An imperative style, high-performance deep learning
  library'',} in \textit{ Advances in Neural Information Processing Systems},
  H.~Wallach {et~al.}, eds., volume~32, p.~8024.
\newblock Curran Associates, Inc., 2019.
\newblock
  \href{http://www.arXiv.org/abs/1912.01703}{\texttt{arXiv:1912.01703}}.

\bibitem{graph_nets}
\href {https://github.com/deepmind/graph_nets}{DeepMind, ``graph\_nets'',}
  2019.
\newblock \url {https://github.com/deepmind/graph_nets}.

\bibitem{tensorflow2015-whitepaper}
M.~Abadi\href {https://www.tensorflow.org}{ {et~al.}, ``{TensorFlow}:
  Large-scale machine learning on heterogeneous systems'',} 2015.
\newblock \url {https://www.tensorflow.org}.

\bibitem{Iiyama:2020wap}
\hrefCMSnoop {}{Y.~Iiyama {et~al.}, ``Distance-weighted graph neural networks
  on {FPGAs} for real-time particle reconstruction in high energy physics'',}
  \href{http://dx.doi.org/10.3389/fdata.2020.598927}{\doi{10.3389/fdata.2020.598927}},
  \href{http://www.arXiv.org/abs/2008.03601}{\texttt{arXiv:2008.03601}}.
  {Accepted by \emph{Front. Big Data}}.

\bibitem{DiGuglielmo:2020eqx}
G.~Di~Guglielmo\hrefCMSnoop {}{ {et~al.}, ``Compressing deep neural networks on
  {FPGAs} to binary and ternary precision with {\texttt{hls4ml}}'',}
  \href{http://dx.doi.org/10.1088/2632-2153/aba042}{\doi{10.1088/2632-2153/aba042}},
  \href{http://www.arXiv.org/abs/2003.06308}{\texttt{arXiv:2003.06308}}.
  Accepted by \emph{Mach. Learn.: Sci. Technol.}

\bibitem{Rankin:2020usv}
\hrefCMSnoop {}{D.~S. Rankin {et~al.}, ``{FPGAs}-as-a-service toolkit
  ({FaaST})'',} in \textit{ 2020 IEEE/ACM International Workshop on
  Heterogeneous High-performance Reconfigurable Computing (H2RC)}.
\newblock 2020.
\newblock
  \href{http://www.arXiv.org/abs/2010.08556}{\texttt{arXiv:2010.08556}}.

\bibitem{Krupa:2020bwg}
\hrefCMSnoop {}{J.~Krupa {et~al.}, ``{GPU} coprocessors as a service for deep
  learning inference in high energy physics'',} (2020).
  \href{http://www.arXiv.org/abs/2007.10359}{\texttt{arXiv:2007.10359}}.
  Submitted to \emph{Mach. Learn.: Sci. Technol.}

\bibitem{Coelho:2020zfu}
C.~N. Coelho\hrefCMSnoop {}{ {et~al.}, ``{Automatic deep heterogeneous
  quantization of deep neural networks for ultra low-area, low-latency
  inference on the edge at particle colliders}'',}
  \href{http://www.arXiv.org/abs/2006.10159}{\texttt{arXiv:2006.10159}}.
  Submitted to \emph{Nat. Mach. Intell.}

\end{thebibliography}\endgroup
\end{document}